\documentclass{article}
\usepackage{emulateapj}
\begin{document}



\newif\ifAMStwofonts

\title{Super-Eddington Fluxes from Thin Accretion Disks?}
\author{Mitchell C. Begelman\altaffilmark{1,}\altaffilmark{2}}
\altaffiltext{1}{JILA, University of Colorado, Boulder, CO 80309-0440, USA; mitch@jila.colorado.edu}
\altaffiltext{2}{Also at Department of Astrophysical and Planetary Sciences, University of Colorado}
       
\date{Accepted 1999.
      Received 1999}
\begin{abstract}
Radiation pressure-dominated accretion disks are predicted to exhibit strong density inhomogeneities on scales much smaller than the disk scale height, due to the nonlinear development of photon bubble instability.  Radiation would escape from such a ``leaky" disk at a rate higher than that predicted by standard accretion disk theory.  The disk scale height is then smaller than that of a similar disk without small-scale inhomogeneities, and the disk can remain geometrically thin even as the flux approaches and exceeds the Eddington limit. An idealized one-zone model for disks with radiation-driven inhomogeneities suggests that the escaping flux could exceed $L_E$ by a factor of up to $\sim 10-100$, depending on the mass of the central object. Such luminous disks would develop strong mass loss, but the resulting decrease in accretion rate would not necessarily prevent the luminosity from exceeding $L_E$.  We suggest that the observed ``ultraluminous X-ray sources" are actually thin, super-Eddington accretion disks orbiting stellar-mass black holes, and need not indicate the existence of a class of intermediate-mass black holes.

\end{abstract}

\section{Introduction}

Magnetized, radiation pressure-dominated atmospheres tend to become highly inhomogeneous on scales much smaller than the radiation pressure scale height, due to the nonlinear development of ``photon bubble instability" (Arons 1992; Gammie 1998).  In this mechanism, gravity squeezes the radiation out of slightly overdense regions, while magnetic tension enhances the overdensity by preventing the gas from spreading sideways. Simultaneously, slightly underdense regions grow more tenuous as radiation force accelerates them; hence the density contrast increases. In a nonlinear analogue of this instability (Begelman 2001; hereafter, Paper I), gas cycles between narrow regions of very high density and much broader regions of very low density.  Gravity and gas pressure dominate the force balance in the dense regions, which resemble gas pressure-supported exponential atmospheres, while the low density regions resemble radiation pressure-driven winds constrained to slide along magnetic field lines.  To complete the cycle, low density matter rejoins the dense phase by passing through an isothermal, gas pressure-dominated shock, while gas moves from the high-density phase to the tenuous phase more gradually, via a slow magnetosonic critical point.  Thus, the inhomogeneities represent nonlinear wave patterns rather than discrete, long-lived filaments.

Thanks to radiative diffusion, the presence of large density contrasts has little effect on the radiation pressure gradient.  Under such conditions the Eddington limit can be circumvented. In the diffusion limit, the radiation flux is inversely proportional to the local density.  Radiation flows readily through tenuous regions while avoiding regions of high density. If the low- and high-density regions are coupled dynamically, the upward radiation force in the tenuous zones can balance the downward gravitational force acting on the dense regions, and {\it global} dynamical equilibrium is maintained. In the steady-state, nonlinear plane-wave solutions presented in Paper I, a combination of ram pressure and magnetic tension forces maintains this coupling. If most of the mass resides in the dense phase while most of the volume is filled with tenuous gas, the total flux through the system can exceed the Eddington limit by a large factor (Shaviv 1998, 2000).
  
This effect should be relevant in luminous accretion disks, the inner regions of which are radiation pressure-dominated (Shakura \& Sunyaev 1973).  If there are no small-scale inhomogeneities, as assumed in the standard theory, the vertical scale height $h$ at a given radius is proportional to the local radiation flux.  If the luminosity approaches the Eddington limit for the central mass, such a disk must inflate to a thickness of order its radius, probably triggering substantial mass loss.  But in the presence of strong density inhomogeneities the disk becomes ``leaky," and more radiation can escape without inflating the disk.  One can even imagine a scenario in which the luminosity exceeds the Eddington limit, while the disk remains geometrically thin.

In this {\it Letter} we analyze the possible effects of nonlinear photon bubble instability on thin accretion disks.  In \S~2 we summarize some generic features of the resulting inhomogeneities in radiation-dominated atmospheres.  In \S~3 we consider their effects on the structure and luminosity of accretion disks, and in \S~4 we discuss the results and propose possible applications to astrophysical systems. In particular, we suggest that the observed ``ultraluminous X-ray sources" are accretion disks radiating at above the Eddington limit for their central masses.

\section{Inhomogenous Radiation-Dominated Atmospheres}

Consider a plane-parallel radiation-dominated atmosphere with column density $\Sigma$ and scale height $h$ in a gravitational field $g(h)$.  The radiation flux passing through the atmosphere is $F$ and the Eddington flux is $F_E (h) = c g(h) / \kappa$, where $\kappa$ is the opacity. The mean mass density is $\bar \rho = \Sigma / 2 h$.  If most of the volume of the atmosphere is filled with tenuous gas with density $\rho_- \ll \bar \rho$, then the condition for overall dynamical equilibrium is (Shaviv 1998)
\begin{equation}
\label{pressureeq}
{p_r \over h} \sim \bar \rho g \approx {\rho_- \kappa F \over c}  ,
\end{equation}
where $p_r$ is the radiation pressure. Since gas cycles between low and high densities via isothermal shocks, there is an additional, local pressure balance condition between the ram pressure of the tenuous phase, $\rho_- v_-^2$, and the postshock pressure, $\rho_+ c_g^2$, in the dense phase, where $c_g$ is the thermal sound speed in the gas.  Downstream of the shock, matter accelerates along each magnetic flux tube (which is assumed to be approximately straight on scales of order the gas pressure scale height, $c_g^2/g \ll h$), and passes through a sonic point where  $v_s = c_g$ (Paper I).  This implies a constant mass flux $\dot m_s = \rho_s c_g$, where $\rho_s$ is the density at the sonic point. If radiative transfer obeys the diffusion law with $F\propto \rho^{-1}$, then  $\rho_s = \bar \rho$.   We therefore have
\begin{equation}
\label{}
\rho_+ c_g^2 \approx  \rho_- v_-^2  = {\dot m^2 \over \rho_-} = {\bar \rho^2 \over \rho_- } c_g^2 .
 \end{equation}
Thus, $\bar \rho$ is the geometric mean between $\rho_-$ and $\rho_+$. 

We solve the two pressure balance equations for $\rho_+$ in terms of $F$ and the other parameters.    These results are verified by detailed calculations of steady-state, plane-parallel inhomogeneities (Paper I). This yields the ``Eddington enhancement factor," $\ell \equiv {F / F_E} \sim {\rho_+ / \bar \rho}$,
giving the factor by which the radiation flux through the atmosphere exceeds that through a homogeneous atmosphere with the same column density and scale height.

The above analysis applies only if the standard radiative diffusion equation is valid {\it locally} throughout the flow, i.e., if the gas is optically thick across any density scale length along the flow, $\rho^2\kappa/ |\nabla \rho | > 1$.  Where this condition is violated the radiation field does not ``see" the change in density and the flux does not adjust to local conditions.   The local diffusion condition is first violated near the minimum density, $\rho_-$ (Paper I), placing a lower limit on the critical mass flux,  $\dot m_s \ga g/ c_g \kappa$.  In this limit, the critical point occurs at a density larger than $\bar \rho$.  The Eddington enhancement factor is therefore 
\begin{equation}
\label{Eddenhance3}
\ell  \equiv {F \over F_E} \sim {\rho_+ \over \bar \rho}\min \left(1, {\bar \rho^2 c_g^4 \kappa^2 \over g^2} \right) .
\end{equation}

To complete the model we specify $\rho_+$ in terms of other atmospheric parameters.  While the nonlinear theory in its current state does not predict the level of inhomogeneity, it clearly indicates an upper limit. If the maximum gas pressure $\rho_+ c_g^2$ exceeds the magnetic pressure, the field will buckle and the crucial dynamical coupling will be lost.   We conjecture that this limit also represents the natural saturation level of inhomogeneous structure.  We therefore set $\rho_+ c_g^2 \sim \xi p_m$, where $p_m$ is the magnetic pressure and $\xi < 1$ is a parameter reflecting the ``inefficiency" of the process by which inhomogeneities grow (e.g., due to the geometry and/or 
time-dependence of the magnetic field, etc.). The value of $\xi$ is not known, but we guess that it might lie in the range $\sim 0.01-0.1$.  In anticipation of applying our model to accretion disks, we set $p_m  \sim \alpha p_r$ where $\alpha$ might also lie in the range $\sim 0.01-0.1$.  Therefore, from eq.~(\ref{Eddenhance3}) we have 
\begin{equation}
\label{ell}
\ell  \sim \xi\alpha {p_r \over \bar p_g} \min\left[ 1, {\bar p_g^2 \kappa^2\over g^2} \right] ,
\end{equation}
where $\bar p_g = \bar \rho c_g^2 \ll p_r$ is the {\it mean} gas pressure.  The corresponding spacing between dense regions --- the ``wavelength" of the inhomogeneities --- is given by $\lambda  \sim \xi\alpha h \min ( 1, {\bar p_g \kappa / g} )$, which is comfortably smaller than the radiation pressure scale height. Note that $c_g$ can be estimated by assuming that the gas is in LTE with the radiation field.  The correct $\alpha$ to use here corresponds to the total magnetic pressure, rather than the $B_r B_\phi$ stress associated with the transport of angular momentum.  Simulations suggest that the latter might be $\sim 10$ times smaller than $p_m$, which is dominated by $B_\phi^2/8\pi$ (Brandenburg et al.~1995; Stone et al.~1996). In applying this model to accretion disks, we correct for the discrepancy by choosing a larger fiducial value of $\xi$, $\xi\sim 0.1$.

\section{Application to Accretion Disks}

Now consider the atmosphere in the context of a thin, Keplerian accretion disk at radius $r$ with accretion rate $\dot M$.  We will primarily consider a one-zone model in which the disk is characterized by a single mean density and scale height; later we will consider the effects of density stratification.  We normalize $r$ to the gravitational radius of a central mass $M$, $\dot M$ to the Eddington accretion rate $\dot M_E = 4\pi GM / \kappa c$, and $h$ to the radius:
\begin{equation}
\label{nondimdisk}
x \equiv {c^2 r\over GM}; \ \  \dot m \equiv {\dot M\over \dot M_E}; \ \ \delta \equiv {h\over r}.
\end{equation}
The local flux due to accretion is then
\begin{equation}
\label{accretionflux}
F = {3 \over 8\pi} {GM\dot M \over r^3}{\cal D} = {3\over 2} {c^5\over GM\kappa} x^{-3} \dot m {\cal D},
\end{equation}
where ${\cal D}\equiv 1- (x_{\rm in}/x)^{1/2}$ for the standard assumption of zero torque at the innermost stable orbit, $x_{\rm in}$ (Shakura \& Sunyaev 1973). The Eddington enhancement factor is 
\begin{equation}
\label{eddenhance}
\ell = {3\over 2} {\dot m {\cal D}\over \delta x } .
\end{equation}
In a standard radiation-dominated accretion disk $\delta$ adjusts so that $\ell = 1$, but this will not be the case in the presence of small-scale inhomogeneities. For an $\alpha-$model viscosity, the dissipated flux is given by 
\begin{equation} 
\label{dissipationflux}
F \approx { 3\over 2} \alpha \Sigma \Omega^3 h^2 = {3\over 2} {c^5\over GM} \alpha \Sigma \delta^2 x^{-5/2}.
\end{equation}
Equating the accretion flux and the dissipated flux we obtain
\begin{equation}
\label{diskdensity}
\Sigma \approx {\dot m {\cal D}\over \alpha \kappa} \delta^{-2} x^{-1/2}; \ \  \bar\rho = {\Sigma \over 2 h} \approx {c^2 \over 2 GM \kappa}{\dot m {\cal D}\over \alpha} \delta^{-3} x^{-3/2} . 
\end{equation}

The radiation and gas pressure are calculated assuming vertical hydrostatic equilibrium and LTE, respectively.  Setting $\kappa = 0.4$ cm$^2$ g$^{-1}$ for electron scattering in ionized hydrogen, and letting $m \equiv  M/M_\odot$, we have
\begin{eqnarray}
\label{diskformulae}
\bar \rho \approx 8 \times 10^{-6} {\dot m {\cal D}\over m \alpha} \delta^{-3} x^{-3/2} {\rm g \ cm}^{-3}; \\
p_r  \approx 8 \times 10^{15} {\dot m {\cal D}\over m \alpha} \delta^{-1} x^{-5/2} {\rm dyn \ cm}^{-2}; \\
\bar p_g  \approx 5 \times 10^{10} \left({\dot m {\cal D}\over m \alpha}\right)^{5/4} \delta^{-13/4} x^{-17/8} {\rm dyn \ cm}^{-2}  . \\
\nonumber\end{eqnarray}
Combining these formulae with equations (\ref{ell}) and (\ref{eddenhance}), we obtain
\begin{eqnarray}
\label{deltaformula}
\delta \sim \max\left[0.3 \xi_{-1}^{-4/13} \alpha_{-2}^{-5/13}(\dot m {\cal D})^{5/13} m^{-1/13}  x^{-5/26} \right. , \nonumber \\ \left. 0.1 \xi_{-1}^{4/21} \alpha_{-2}^{-5/21}(\dot m{\cal D})^{5/21} m^{-1/21}   x^{1/14}  \right],
\end{eqnarray}
where we have normalized $\xi$ to 0.1 and $\alpha$ to 0.01. A similar equation for $\ell$ can be derived from eq.~(\ref{deltaformula}), using eq.~(\ref{eddenhance}).  Inhomogeneities affect the disk structure only if $\ell > 1$, implying
\begin{eqnarray}
\label{xinhom}
x < x_{\rm inhom} \sim \min\left[7 \xi_{-1}^{8/21} \alpha_{-2}^{10/21}\dot m^{16/21} m^{2/21}, \right. \nonumber \\
\left. 9 \xi_{-1}^{-8/45} \alpha_{-2}^{2/9}\dot m^{32/45} m^{2/45} \right] ,
\end{eqnarray}    
where we have set ${\cal D} \sim 1$ for $x_{\rm inhom} \gg x_{\rm in}$.

Now consider the maximum possible mass flux that can be carried by the disk. Since inhomogeneities are ineffective outside $x_{\rm inhom}$, we have $\dot m_{\rm max}(x_{\rm inhom}) \sim 2 x_{\rm inhom} /3 $, corresponding to a thick disk [$\delta (x_{\rm inhom})\sim 1$] with a local radiation flux equal to the Eddington limit. Any excess matter supplied at larger $x$ presumably will be driven away by radiation pressure before reaching $x_{\rm inhom}$.  For $\dot m$ close to $\dot m_{\rm max}$ and $\xi \la 0.1$, local radiative diffusion should apply through most of the inhomogeneous region, and certainly in the inner regions where most energy is released.  We therefore have
$\dot m_{\rm max} (x_{\rm inhom}) \sim  580 \xi_{-1}^{8/5} \alpha_{-2}^2 m^{2/5}$.
By setting $\delta (x) \sim 1$ at $x < x_{\rm inhom}$, we find that the carrying capacity of the disk {\it decreases} with decreasing $r$, $\dot m_{\rm max} \propto x^{1/2}/ {\cal D}$, down to the radius $x \sim 4 x_{\rm in}$ at which $x^{1/2}/ {\cal D}$ has a local minimum. Thus, the mass flux reaching the central object may be much smaller than that entering the inhomogeneous region, $\dot m_{\rm in, max} \sim 4 (x_{\rm in} / x_{\rm inhom})^{1/2}\dot m_{\rm max}(x_{\rm inhom})$. Nevertheless, the {\it luminosity} can exceed $L_E$.

According to the one-zone model, the maximum possible disk luminosity satisfies  
\begin{equation}
\label{LoverLE}
{L_{\rm max} \over L_E} > \epsilon \dot m_{\rm in, max} \sim 30 \xi_{-1}^{4/5} \alpha_{-2} \left({m\over 10}\right)^{1/5}\left({\epsilon \over 0.1}\right)\left({x_{\rm in}\over 6}\right)^{1/2} , 
\end{equation}
where $\epsilon$ is the accretion efficiency and the inequality reflects the fact that $\dot m$ increases with $r$.  However, mass loss from the disk's upper layers may limit the maximum attainable luminosity. Such a wind must arise, in a super-Eddington disk, from strata where the mean optical depth is so small that density inhomogeneities provide insufficient modulation of the flux.  If the radiation becomes ``trapped" in the accelerating outflow it will drive mass loss at a rate comparable to or exceeding the inflow rate, possibly regulating the accretion luminosity to a value $\sim L_E$. But if the gas remains sufficiently inhomogeneous in the acceleration zone, radiative losses due to diffusion could reduce the energy supply available for unbinding the gas. The mass loss rate would then be lower (Shaviv 2001). Mass loss could be further inhibited if dynamo processes in the disk maintain a sufficiently dense array of closed magnetic flux loops in a magnetically dominated corona (Miller \& Stone 2000).  If $\ell < p_m \kappa / g$, then gas propelled upward into closed field lines anchored in the disk can become trapped by the magnetic tension.  Once its optical depth reaches $\sim \ell$, a trapped blob begins to be weighed down by gravity, and if  $\tau$ reaches $\sim p_m \kappa / g$ before the field opens up, the gas will overcome the tension and drop back onto the disk.   

\section{Discussion}

We have argued that the radiation-dominated, inner regions of accretion disks may be able to approach or even exceed the Eddington limit, without becoming geometrically thick or suffering catastrophic mass loss.  The disks become highly inhomogeneous on small scales, and correspondingly porous to radiation leakage, due to the nonlinear development of photon bubble instability (Gammie 1998; Paper I). The possible enhancement in radiation flux is limited by the strength of the magnetic field, which in turn is related to viscous dissipation and angular momentum transport via the magnetorotational instability (MRI, which operates concurrently with photon bubble instability [Blaes \& Socrates 2001; Turner, Stone, \& Sano 2001]). Greater porosity requires larger density constrasts, which place greater stresses on the magnetic field. Beyond a certain flux, the magnetic tension will be overwhelmed and the field will buckle. Simple one-zone models suggest that accretion disks around stellar-mass black holes could radiate as much as $\sim 10$ times the Eddington limit.  A weak mass dependence implies that disks around supermassive black holes could exceed the Eddington limit by an even larger factor, perhaps $\sim 100$ or more.  

Although the magnetic configurations in accretion disks are highly dynamical and do not resemble the quiescent, uniform fields assumed in our analytic models, this time-dependence need not impede the development of inhomogeneities.  The linear growth time for photon bubble instability is $\sim (\lambda / g)^{1/2}$, the free-fall time across the wavelength (Gammie 1998).  This is shorter than the orbital time, which determines the evolution timescale of magnetic structures due to MRI (Blaes \& Socrates 2001).   

Catastrophic mass loss from the disk's upper layers poses the most serious threat to the production of super-Eddington luminosities.  If the luminosity is effectively trapped in the outflowing gas it could deplete the accretion flow, regulating the maximum luminosity to $\sim L_E$. But if inhomogeneities persist into the acceleration zone the mass loss could be lower, leaving the accretion flow largely intact. A magnetically structured corona with a sufficient number of closed field lines could also inhibit mass loss by trapping and recycling some of the expelled gas.  Both speculations will have to be checked by detailed calculations, including numerical simulations. 

Fluxes $\ga 10 L_E$ could explain the existence of ``ultraluminous X-ray sources" (ULXs: Fabbiano 1989; Colbert \& Mushotzky 1999; Makishima et al.~2000; Fabbiano, Zezas, \& Murray 2001).  The high luminosities of these sources have led to suggestions that they represent a new class of intermediate-mass black holes.  However, the observed luminosities are also explainable in terms super-Eddington fluxes from disks around black holes of a few solar masses.  This is consistent with anomalies reported by Makishima et al.~(2000), who performed ``multicolor blackbody disk" fits to ULX spectra and found that the best-fitting models were considerably hotter and more compact than expected for $\sim 100 M_\odot$ black holes. King et al. (2001) have also proposed that ULXs are sources producing super-Eddington fluxes, but they posit that the intense radiation is beamed within a small solid angle.  In contrast to our proposal, which predicts a roughly isotropic flux, they require a large population of misdirected sources. However, we agree with them that these systems should not be very long-lived, since an object accreting at $\sim 10$ times the Eddington limit would double its mass in a few million years.  

We make two other predictions based on the analysis presented in this paper. First, inhomogeneous accretion disks radiating above the Eddington limit should possess ``cold" winds driven by continuum radiation pressure. In contrast to thermally driven outflows, these winds should be at or below the Compton temperature of the accretion disk radiation, and well below the virial temperature.  They should have substantial optical depths and may therefore modify the underlying disk radiation.  Moreover, if ``hot" flares (energized by magnetic reconnection, for example) percolate through such a wind, Compton downscattering of the hard radiation could produce a prominent ``soft excess".  It may be difficult to distinguish such winds from large-scale hydromagnetic flows (Blandford \& Payne 1982; K\"onigl \& Kartje 1994), if the latter also possess large optical depths. 

Second, there may be a population of ``ultraluminous" AGNs, i.e., supermassive black holes radiating at well above their Eddington limits. Candidates could be identified by looking for discrepancies between the black hole mass as estimated from the properties of the host galaxy (Ferrarese \& Merritt 2000; Gebhardt et al. 2000) or by reverberation mapping and line widths, and the lower bound on the mass according to the Eddington limit.  An epoch of super-Eddington accretion at high $z$ could lead to the rapid growth of supermassive black holes by accretion.

Finally, we note that SS 433 may be an example of an object which is accreting at a rate that far exceeds even the upper limit attainable in a ``leaky"  disk (King, Taam, \& Begelman 2000).  Most of the matter initially in the accretion flow is therefore blown away before it reaches the vicinity of the central object. Nevertheless, our analysis suggests that the mass flux reaching the central object could exceed the Eddington limit by a substantial factor. Much of the energy released by the accreting matter would be available for driving an outflow and perhaps goes into powering the $0.26c$ jets, which are believed to carry a super-Eddington energy flux (Begelman et al.~1980).  We presumably do not see a highly super-Eddington radiative flux because of the large optical depth of intervening matter.

\section*{Acknowledgments} 
This work was supported in part by NSF grant AST-9876887. I am grateful to Neal Turner, Roger Blandford, and the referee, Nir Shaviv, for helpful comments.

\label{lastpage}
\end{document}